\documentclass[aps,prl,showpacs,twocolumn,superscriptaddress,amsmath,amssymb,floatfix]{revtex4}
\usepackage{tabularx}
\usepackage{bm}
\usepackage{epsfig,subfigure}
\usepackage{multirow}
\usepackage{graphicx}
\usepackage{color}
\usepackage{amsfonts}
\usepackage{exscale}
\usepackage{amsbsy}
\usepackage[colorlinks,urlcolor=blue,linkcolor=blue,anchorcolor=blue,citecolor=blue]{hyperref}

\begin{document}

\title{Combining Grassmann algebra with entanglement renormalization method}

\author{Jie Lou}
\affiliation{Department of Physics and State Key Laboratory of Surface Physics, Fudan University, Shanghai 200433, China}
\affiliation{Collaborative Innovation Center of Advanced Microstructures, Nanjing 210093, China}
\author{Yan Chen}
\affiliation{Department of Physics and State Key Laboratory of Surface Physics, Fudan University, Shanghai 200433, China}
\affiliation{Collaborative Innovation Center of Advanced Microstructures, Nanjing 210093, China}

\begin{abstract}
By combining the Grassmann algebra with multi-scale entanglement renormalization
ansatz (MERA), we introduce a new unbiased and effective numerical method for
simulating 2D strongly correlated electronic systems.
The new GMERA method inherits all the advantages of MERA, which constructs the
variational wave function based on complicated tensor network.
Besides it can deal with fermionic properties of the system due to Grassmann algebra
through local tensor contractions.
This general method can treat different tensor network structures in a universal way.
We show several benchmark calculations of the GMERA method, including the free fermion model,
tight binding model, as well as the $t$-$J$ model with hole doping.
\end{abstract}

\date{\today}


\maketitle
\section{I. Introduction}
The past three decades witnessed the rapid progress of synthesis and characterization of a large amount of strongly correlated electronic
materials.\cite{Anderson,PALee}
The nature of two-dimensional strongly correlated electron systems remains one of the biggest challenges in condensed matter physics
for both theorists and experimentalists.
Various theoretical models including electronic models as well as quantum spin models
have been proposed based on these materials, all featuring strong correlations and quantum fluctuations.
Analytic solution to these models are still lacking, and
numerical methods become efficient approaches to unveil their physical nature.
Many successful sophisticated numerical methods have been proposed. Among them, the
quantum Monte Carlo (QMC) method is extremely powerful which can
describe large system to very high accuracy.
Unfortunately, QMC suffers from ``sign problem'', which makes it difficult to
be applied in systems with frustrations or in presence of fermions.\cite{QMC}
On the other hand, a group of variational methods based on matrix product
state ansatz (MPS) or tensor product ansatz (TPS) have been developed during the past years, such as the
density matrix renormailzation group (DMRG),\cite{DMRG1,DMRG2}
projected entanglement pair states (PEPS),\cite{PEPS}  the multiscale entanglement renormalization ansatz (MERA) {\it etc}.\cite{MERA1}\cite{MERA2}
These methods are targeted to describe systems with long range entanglement.
There are numerous successful applications of these MPS/TPS based techniques to the study of
spin liquid states in frustrated quantum spin systems, which are
known for harboring long range entanglement and huge degeneracy of low-lying excited states.

There has been quite a few attempts to apply MPS/TPS based methods to fermionic system.
\cite{TPS_f}\cite{MERA_f}
These pioneer methods consider the tensor network as fermion operator circuits,
and usually involve complicated swaping scheme for crossed fermion operators in the circuits.
These method are capable of dealing with relatively simple tensor networks,
such as that of PEPS, or 1D MERA.
However, when the network becomes rather complicated, for instance, multi-level tensor networks for 2D system,
 the operator circuits appear to be extremely complicated and very difficult and
tedious to cope with. Recently, the Grassmann tensor product state (GTPS) method has been proposed.\cite{GTPS}\cite{GTPS_b}
It has been further shown that based on the fermion coherent state representation,
all standard numerical method based on TPS ansatz can be generalized into GTPS.
In the fermion coherent state picture,
a single Grassmann variable is used to indicate fermion/boson
nature of each channel of inner index on virtual bonds, hence significantly
simplify the representation in numerical calculation.
Furthermore, reordering of Grassmann variables during tensor contractions
recover anti-commutating relation of fermions.
The key advantage here is that this procedure only involve local tensors and their
associated Grassmann number, hence dramatically reduce the complexity of the computation.
The GTPS method has been successfully applied to the $t$-$J$ model on the honeycomb lattice.\cite{GTPS_tj}

In this paper we develop a new efficient GMERA method to simulate fermionic system,
by combining the Grassmann algebra with the MERA method.
The GMERA method inherits all the advantages of MERA, which constructs the
variational wave function based on complicated tensor network.
The implementation of Grassmann algebra here allows us to obtain fermion statistics
through local tensor contractions, hence significantly reduce the complexity of the
problem when complicated MERA tensor network structure is involved.
We show several benchmark calculations of the GMERA method, including the free fermion model,
tight binding model, as well as the $t$-$J$ model with hole doping.
Our GMERA method may serve as an alternative unbiased and effective numerical method for
simulating 2D strongly correlated electronic systems.

\begin{figure}
\centerline{\includegraphics[angle=0,width=8.2cm]{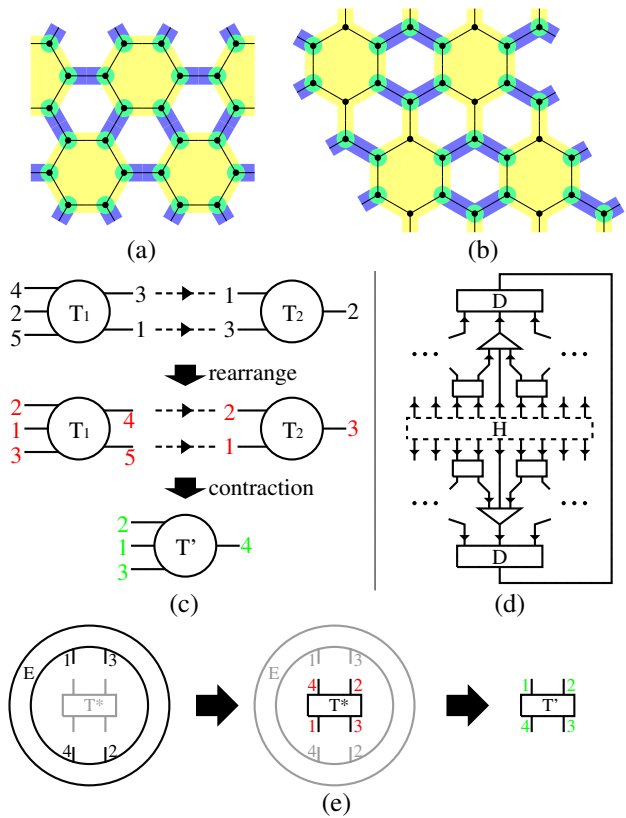}}
\caption{(Color online) Upper two panels: tensor network structures used
in our benchmark calculation for the $t$-$J$ model on honeycomb lattice.
Blue region covers lattice sites involved in local disentangler tensors,
and yellow region shows sites coarse-grained in isometry tensors.
Lower panels: (c) the rearrangement of tensor indices (and associated Grassmann
numbers) after which two tensors can be contracted sign-free as bosonic counterparts;
(d) the optimization of individual tensor;
(e) direction (stands for order of Grassmann numbers $d\theta_i d\theta_j$)
on each links in tensor networks.
}
\label{illustration}
\end{figure}

The remainder of the paper is organized as follows. In Sec. II, we introduce the main methodology of GMERA.
The method constructs the variational wave function based on complicated tensor network.
Meanwhile it can deal with fermionic properties of the system due to Grassmann algebra
through local tensor contractions.
In Sec. III, we present three benchmark calculation results, including the free fermion model,
tight binding model, as well as the $t$-$J$ model with hole doping. We make comparisons of ground-state energy
with exact diagonalization calculation as well as DMRG results and a very good agreement is achieved.
Finally a brief summary is provided.

\section{II. Methodology}

Before discussing the details of GMERA method, we would like to comment a few different facts between
MERA and other TPS based methods (PEPS, TERG {\it etc.}).
In standard TPS,
the tensor network need to be contracted to optimize the variational wave function,
or measure physical quantities.
As a consequence, the computation cost rises up very fast due to the contraction operations
of tensor network in two or higher dimensions,
as the number of open links of each bulk of the lattice increases with its size.
This is a simple reflection of the area law in 2D.
We have to make compromise and various approximations have been adopted to control the cost.
For instance, cut off of virtual index for each link is introduced,
only those most significant element is kept through singular value decomposition (SVD).

The MERA method is a slightly different version of TPS, since the tensor network
structure resembles a tree tensor network which coarse grain the system,
and its design is based on the concept of entanglement renormalization.
The key idea is to introduce unitary transformations by local tensors named disentangler,
removing short range entanglement before coarse-graining step.
The ultimate goal is to avoid accumulation of short range entanglement under multiple RG processes, so that
critical behavior of arbitrary large system can be addressed.
On the other hand, one can simply apply single or two layers of RG process,
to appreciate the effect of disentangler, so that relatively long range entanglements
of a finite system can be captured through limited degrees of freedom in the coarse grained system,
by reasonable computational cost.
Several successful attempts to study frustrated quantum antiferromagnets have been achieved.
\cite{kagome}\cite{triangular}\cite{shastry}

We have to point out that in systems with large entanglements such as spin liquid state,
the local disentangler is not capable to remove accumulation of entanglement completely in the
RG process. As a result, the degrees of freedom we used to describe coarse grained system greatly affect
the accuracy of MERA result.
Whether MERA method is more cost-efficient than PEPS/TERG is questionable,
but it does have several advantages.
For instance, MERA is an exact variational method and no approximation (cutoff)
is made in the MERA wave function. As a result, MERA result provides upper-bound of the ground state energy.
In addition, MERA can easily deal with any kind of periodic or open boundary conditions versatilely,
unlike in certain TPS based methods where different boundary conditions may be tricky to be dealt with.

In principle, the GMERA method is very similar to other GTPS approach, only with a few complications.
The key point remain the same: we introduce fermion coherent state representation into the system,
and treat physical indices as a Grassmann variable.
For each virtual links between two tensors, degrees of freedoms are spitted into boson/fermion
channels, with a Grassmann number ($0$ for boson and $1$ for fermion) assigned to each channel.
Due to the fact that Grassmann number follows same anti-commutation rule as fermions,
each tensor must have even fermion parity.
As a result, all tensors (with Grassmann variable attached) can be freely exchanged during contraction
without any sign problem, and contraction of the full tensor network can be performed in the
same way as in fermion-free situation.
The fermion statistic is recovered during contraction of two tensors
where a Grassmann integration has to be performed, which requires reordering of Grassmann numbers
who share same anti-commutation rule as fermions, as shown in Fig.~\ref{illustration} (c).
It is obvious here that GTPS class of methods hold one special advantage over previous attempt
of TPS method for fermion system, namely, exchange of Grassmann number (equivalent to fermion
operation) can be treated locally and universally, without complication from neither lattice
nor tensor network structure.

Notice that we have parity conservation for each tensor in the GTPS scheme.
Namely, the total parity of fermions (1 for even number of fermions
and -1 for odd) must be even.
It is quite natural to expect, as fermions are always created in pairs.
In the case of GMERA, this holds true as well,
and can be simply understood as the following: fermion parity is conserved
during all operations, including coarse graining (isometry) and unitary transformation (disentangler).
Notice that the total parity of the coarse grained system can be even/odd,
which corresponds to states in original system with even/odd number of fermions, respectively.
This does not fix number of fermions in the system.
On the contrary, we want to point out that the GMERA method adopts the grand canonical ensemble.
If pair correlation/annihilation term $c^{\dagger}c^{\dagger}+h.c.$ exists,
the ground state can be superposition of wave functions
hosting even/odd number of fermions,{\it, e.g.},
$|\Psi \rangle = |0 \rangle + |2 \rangle + |4 \rangle$....
In practice, the number of fermions in the system does not depend on how our calculation
being performed.
This feature is quite different from the ED or the DMRG method, which normally project the
wave function into fixed fermion number sub-Hilbert space.

The major difference between (G)MERA and other (G)TPS method is the way how tensors are optimized.
An illustration can be found in Fig.~\ref{illustration} (d).
In the case of MERA, the whole tensor network is contracted, leave only one tensor T that
is to be optimized.
The resulted tensor E is the environment of the target tensor, and SVD is then applied
to achieve the new optimized tensor T$'$.
This procedure has to be performed for each tensor one after another.
There are several subtle points here when Grassmann numbers are introduced into the problem.
First of all, in the GTPS scheme including GMERA, the order of all indices of any tensor is crucial.
switch indices will result in fermion sign since their associated Grassmann numbers are shuffled
at the same time.
As a result, it is wise to prepare a preset order of indices for each tensor,
and make sure each time all indices of the tensor is converted back to the conventional
order before stored during simulation.
It is especially important to keep this in mind during optimization of tensors,
as the new tensor T* we obtain after SVD of the environment E has all indices reversely arranged
(so that the SVD is sign-free).
This tensor need its indices and associated Grassmann numbers rearranged to convert
to its conventional setup T$'$.

Another important point is that each link between two tensors is assigned a  ``direction'',
which notates the order of associated Grassmann numbers $d\theta_1,d\theta_2$
that should be written down during Grassmann contraction process.
Reversion of the direction will induce a negative sign if $d\theta_1=d\theta_2=1$.
The direction can be chosen arbitrarily before simulation, but be kept the same afterwards.
One should be cautious when treating these ``directions'' of optimized tensor T$'$.
Keep in mind that they should be flipped into the preset directions as well.
In practice, for simplicity, we choose the ``direction'' of all links pointing from original lattice to
the coarse grained lattice, as shown in Fig.~\ref{illustration} (e).

\begin{figure}
\centerline{\includegraphics[angle=0,width=7.8cm]{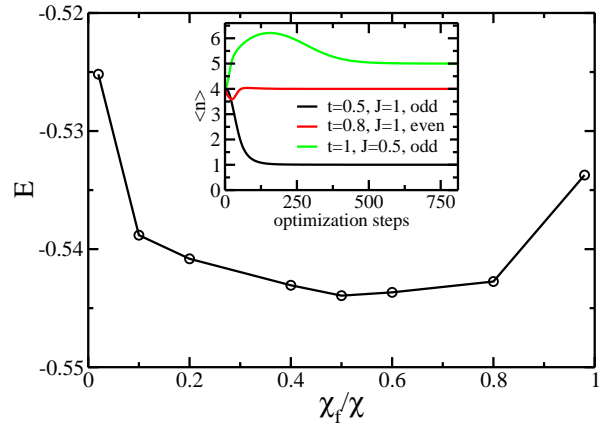}}
\caption{The ground state energy we obtained from variational calculation
depends on the ratio between the numbers of fermionic/bosonic channel, with fixed virtual bond dimension $\chi$.
Here we show energies of such optimized states as a function of the ratio of fermionic channels.
The calculation is for the $t$-$J$ model on a $12$-sites lattice with single hole doping, at $t/J=0.5$.
Inner panel: evolution of number of holes $\langle n \rangle$ as function of
optimization steps.
}
\label{ratio}
\end{figure}
Another distinctive feature of GMERA is how the total degrees of freedom
(virtual bond dimension $\chi$) on each link is splitted into fermion/boson channels.
For standard GTPS method, the number of boson/fermion channels is
automatically determined by SVD, and those channels with most significant value is kept.
In the GMERA case, the number of boson/fermion channel is set before the simulation.
At first glance, this just create additional task that one has to find the optimum
ratio through some trail runs before commit to a full calculation.
In reality, we can have pretty good guess of the correct ratio in most case,
and we can expand number of boson/fermion channels at any time in simulation easily.
As one example, in our calculation of $t$-$J$ model (which we will elaborate to explain later),
we find that half-half distribution of bosonic/fermionic channels is best suited to obtain
the ground state, as shown in Fig.~\ref{ratio}.

\begin{figure}
\centerline{\includegraphics[angle=0,width=7.8cm]{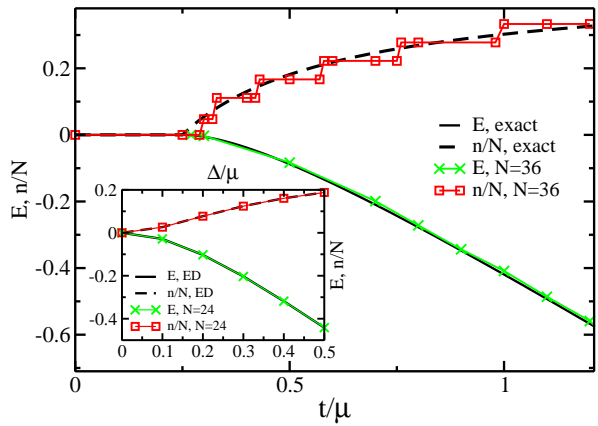}}
\caption{Results of the tight binding model on the square
lattice with $6 \times 6$ sites.
$\mu$ is set to $1$, and pairing interaction $\Delta$ is tuned.
Our results are consistent with analytical solution (in thermodunamicl limit)
shown as black lines in the background.
Inner panel: Results of the free fermion model on the hexagon lattice
with $24$ sites.
Black solid/dashed lines show results from exact diagonalization.
}
\label{free_fermion}
\end{figure}
Furthermore, we want to emphasis a few technical aspects of the GMERA method.
First of all, tensors network in GMERA has much more complicated structure than those in other
GTPS method ({\it i.e.}, GTERG, GPEPS). Tensors (especially isometries) tends to have quite a number
of indices, not to mention those large tensors we obtain during contraction of the network.
As a result, reordering of all Grassmann indices can be quite a formidable obstacle in the
sense of computational cost. There are two essential techniques to solve this problem.
Firstly, before contraction of two tensors A and B, we can rearrange their indices
individually, and make sure after rearranging, they can be contracted sign-free (as boson tensors).
This is much easier job than performing Grassmann integration during the contraction.
Secondly, we can always record the effect of rearranging for individual tensor
(A for example) by saving a bit-tensor $A_b$, whose elements are nothing but fermion signs which
the corresponding element in A obtained through rearranging.
These bit-tensors are not too large, and manageable to save in the memory.
With these information directly accessible in following simulation,
the GMERA runs at the speed similar to the bosonic code (at worst a few times slower).
\begin{figure}
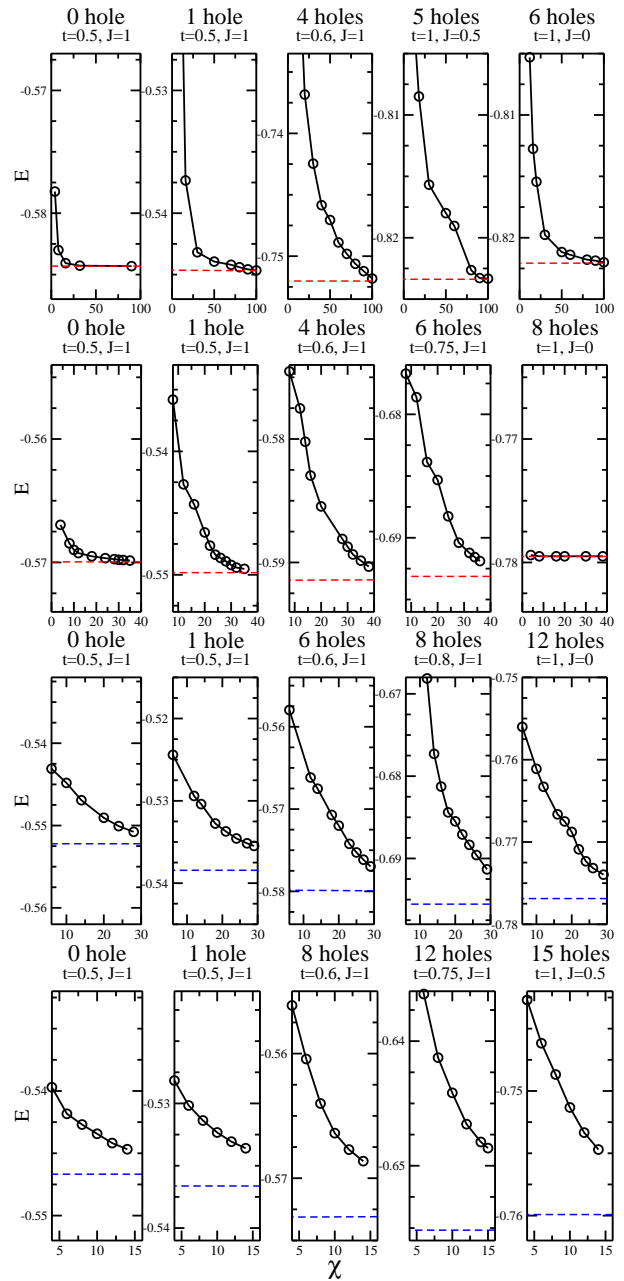

\centerline{\includegraphics[width=8cm,clip]{converge_o.eps}}
\centerline{\includegraphics[width=8cm,clip]{converge_u.eps}}
\centerline{\includegraphics[width=8cm,clip]{converge_j.eps}}
\centerline{\includegraphics[width=8cm,clip]{converge_v.eps}}
\caption{Ground state energy convergence from GMERA calculation. Horizontal axis $\chi$ stands for the virtual
bond dimension for coarse grained system, which determines the accuracy of the calculation.
The four panels from top to bottom represent the different clusters with
$N=12,16,24$ and $32$ sites for various hole doping.
Red and blue dashed lines stands for the ground state energy obtained from exact diagonalization and DMRG results, respectively.
}
\label{energy_compare}
\end{figure}

\section{III. Benchmark Calculations}
We performed several benchmark calculations on various models and lattices to test GMERA method.
As a simple starting point, we calculate the spinless free fermion model on the honeycomb lattice.
The Hamitonian is of the form
\begin{eqnarray}
H=-\Delta \sum (c^{\dagger}_ic^{\dagger}_j + h.c.) +\mu \sum c^{\dagger}_i c_i,
\end{eqnarray}
where $\mu$ is the chemical potential which controls number of fermions in the system.
When $\mu$ is fixed to $1$, and $\Delta$ is switched on, we may create more pairs of fermions.
Our numerical results are shown as the inset in Fig.~\ref{free_fermion}.
This is a very simple model and our result is perfectly consistent with that obtained from
exact diagonalization, with energy difference less than $0.1\%$.

The second benchmark calculation is for the tight binding model on the square lattice. 
The Hamiltonian can be written as
\begin{eqnarray}
H=-t \sum (c^{\dagger}_ic_j + h.c.) +\mu \sum c^{\dagger}_i c_i,
\end{eqnarray}
and the exact solution of this simple model is well-known.
Since the system is of grand canonical ensemble, we have more fermions by increasing $t$.
We compare our calculation results obtained from $N=36$ sites lattice
to the exact solution (in thermodynamic limit) and quantitative agreement has been made, as shown in Fig.~\ref{free_fermion}.
Notice that our system has finite size, and we can only have integer number of
fermions in the system, so the filling factor $n/N$ will naturally show plateau like behaviors.
In the present calculation, we use a simple MERA tensor structure designed
for the square lattice proposed by Vidal {\it et.al.}\cite{MERA1}.

The third model we study is the well-known $t$-$J$ model, which describes a doped Mott insulator.  
Its model Hamiltonian can be written as
\begin{eqnarray}
H=-t \sum_{\langle i,j \rangle,\sigma} (\tilde{c}^{\dagger}_{i,\sigma}\tilde{c}_{j,\sigma} + h.c.) +J \sum_{\langle i,j \rangle} {\bf S}_i \cdot {\bf S}_j +\mu \sum_{i,\sigma} n_{i,\sigma},
\end{eqnarray}
where $n_{i,\sigma}=\tilde{c}^{\dagger}_{i,\sigma}\tilde{c}_{i,\sigma}$ and the term $-\frac{1}{4}n_in_j$ is ignored.
We start from the half-filling limit (AF Mott insulator), and dope the system with holes.
For simplicity we use the holon representation, in which we treat spins as bosons,
and holes as fermions.
The creation operator $\tilde{c}^{\dagger}_{i,\sigma}$ can be decomposed as $h^{\dagger}_ib_{i,\sigma}$,
where $h$ and $b$ are creation operators of holon and spinon,
and the hopping term hamiltonian can be rewritten as
\begin{eqnarray}
H_{hopping}=-t\sum_{\langle ij \rangle, \sigma}h^{\dagger}_jh_ib^{\dagger}_{i,\sigma}b_{j,\sigma}.
\end{eqnarray}

Notice that we have the no-double-occupancy constraint, as a result,
the complete basis for each site consists of three states:
spin up/down $|\uparrow (\downarrow) \rangle = b^{\dagger}_{\uparrow ( \downarrow)} | 0 \rangle$
and the hole state $|o\rangle =h^{\dagger}_i|0\rangle$.
In practice, in definition of the Hamiltonian in GMERA method,
we assign two spin states to bosonic channels of each site's physical indices,
and the hole state $|o\rangle$
to the remaining fermionic channel.

We want to point out here that due to grand canonical ensemble,
the number of fermions (holes) in the $t$-$J$ model is determined by
the ratio $t/J$, by optimizing the ground state energy.
Alternatively, given $t/J$,
we can achieve desired doping by tuning chemical potential $\mu$.
Notice that for the $t$-$J$ model,
in absence of pair correlation/annihilation, the number of fermions in the
system should converge to integer number, as shown in right panel of Fig.~\ref{ratio}.

\begin{figure}
\centerline{\includegraphics[angle=0,width=8.0cm]{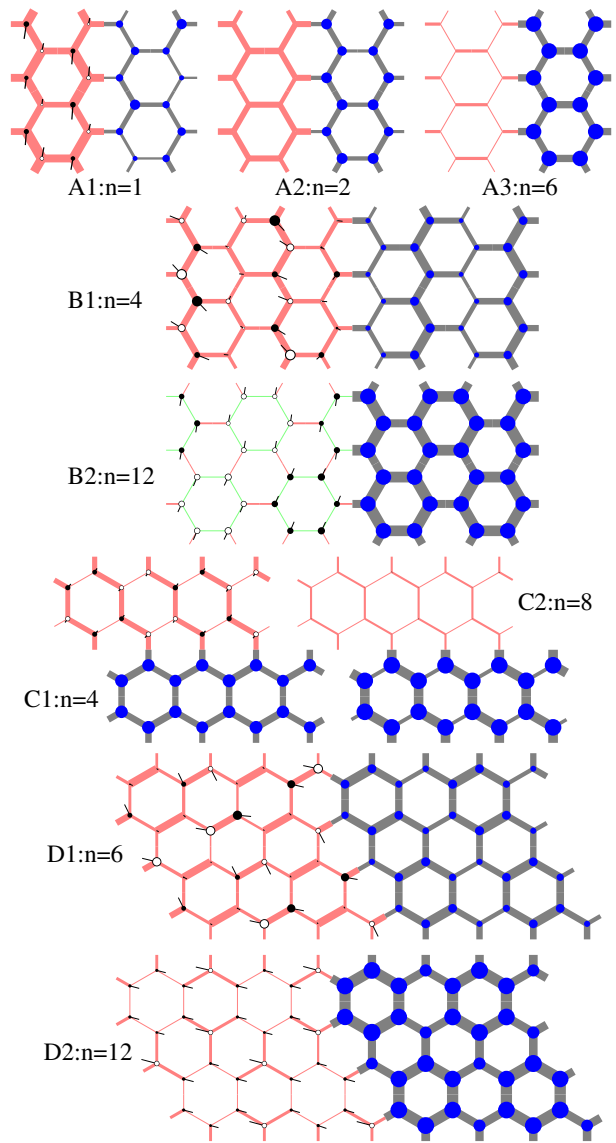}}
\caption{Local physical quantities measured in the ground state of the $t$-$J$  model.
Panels labelled A,B,C,D shows results for lattices with $12$,$24$,$16$ and $32$ sites, respectively.
Each configuration consist of two unit cells, with one illustrating spin related quantities,
and another showing hole related ones.
The open/closed circle and a vector plotted on each site of the left(up) unit cell
represents spin components $\langle S_z \rangle$ and $\langle S_x \rangle,\langle S_y \rangle$.
Red/Green bonds represent AF/ferromagnetic spin correlations $\langle {\bf S}_i \cdot {\bf S}_j \rangle$,
with their width proportional to the strength.
In right/down unit cells, The blue dots shows the
hole occupation number $\langle n \rangle$, and the width of each shaded bonds
represent the strength of hopping term between two sites.
}
\label{configuration}
\end{figure}
The nature of the ground state of the $t$-$J$ model is still controversial,
reflecting the complexity of the quantum many-body problem as well as the lacking of effective computational and analytical tools
for treating strongly correlated electronic systems.
The goal of our benchmark calculation here is not to fully solve the problem;
instead, we compare our result with exact diagonalizations for small systems
to establish the validity of our method. Furthermore, we
compare our result for larger system with DMRG calculations, and show that GMERA is also capable of dealing
with larger lattice sites effectively.

We study the honeycomb lattice with four different sizes $12$, $24$, $16$, $32$,
and two types of periodic boundary conditions.
The tensor structure we use in these calculations consists of one layers of isometry tensors
and one layer of disentanglers.
In calculations of $N=12,24$ lattices, the isometry tensor coarse grain 6 sites into one,
whereas in calculations of remaining two systems, the isometry tensor covers 8 sites.
Details of tensor network structures
are shown in top panels of Fig.~\ref{illustration}, and the shapes of unitcell
can be found in Fig.~\ref{configuration}.

First of all, to verify the validity of GMERA method,
we compare the ground state energy we obtain from GMERA calculation to exact diagonalization,
as shown in Fig.~\ref{energy_compare}.
We fix $\mu=0$ and switch on $t/J$. By increasing the strength of hopping term,
the system in ground state may correspond to more holes.
It is clear that for small system $N=12$ and $N=16$, the result of GMERA is perfectly consistent
with ED.
Notice that GMERA method is a rigorous variational method with no approximation;
as a result, the ground state we obtain is exactly equal to the true ground state if
the accuracy is sufficient high. No approximation is introduced in the process.
It is worthwhile to mention that the ground state energy converges very fast when
number of holes in the system is small.
When more holes are introduced, the convergence becomes much slower, require larger dimension 
$\chi$ to achieve same level of accuracy.
This is due to the fact that fermions introduce additional entanglements in the system,
and similar feature was discussed.\cite{MERA_f} 
An exception is the case of $n=N/2$, namely the system is half doped with holes.
Here we choose $t=1, J=0$, which convert the $t-J$ model to the spinless tight binding model.
Notice the peculiar situation for the $16$ sites lattice with $8$ holes.
The ground state energy converges extremely fast with very small $\chi$,
indicating near-zero long range entanglement of the ground state.
The reason of this puzzling phenomena is not clear.

For larger system, the energy converges smoothly, although the virtual bond dimension $\chi$ is
not sufficiently large, and our ground state energy
can not compete with high accuracy of DMRG method.
Even though we still have not fully explored the potential of GMERA method, the agreement between our result
 and DMRG method is good enough. For $N=24$ lattice, the energy difference is less than $0.5\%$, while in the $N=32$ case,
the difference can be as large as $0.7\%$.
We would like to point out that no symmetry has been implemented in our GMERA algorithm yet,
and our code is roughly optimized. In principle, we should be able to push the accuracy of our calculation much higher,
and we will leave this for future studies.

We measure all local physical quantities of the ground state obtained in our simulation,
including spin component $S^{x}_i,S^{y}_i,S^{z}_i$, hole occupation number $\langle n \rangle$,
spin correlations $\langle {\bf S}_i \cdot {\bf S}_j \rangle$, as well as nearest neighbor hopping
$\sum_{\sigma} \tilde{c}^{\dagger}_{i,\sigma}\tilde{c}_{j,\sigma}$.
Results are plotted in Fig.~\ref{configuration}.
Here we fix $t/J=3$, and dope the system with different number of holes by tuning
chemical potential $\mu$.
It is worth to point out that for small systems, the local quantity configurations
we obtain and are fully consistent with ED.

\begin{figure}
\centerline{\includegraphics[angle=0,width=5.9cm]{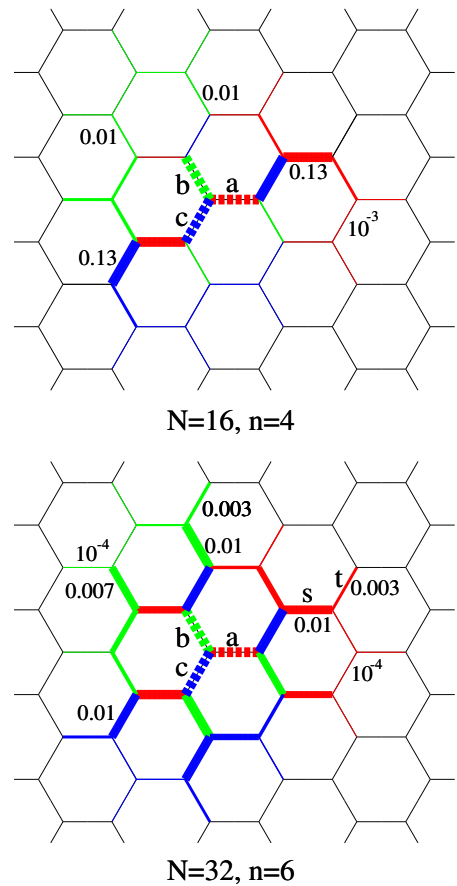}}
\caption{Distribution of pair correlation function $C_p(i,j)$.
Upper and lower panels shows results from $N=16$ and $N=32$ system, respectively.
In both cases, $t/J=3$ and systems are adequately doped so that superconductivety emerges.
The dashed bonds shows the position of three reference bonds $(i, i+{\bf e}_{x,y,z})$, notated
as $a$,$b$ and $c$.
For each solid bond at position $(j, j+{\bf e})$, its width represents the amplitude
of pair correlation $C_p(i,j)$,
and the color denotes the bond $(i, i+{\bf e})$ it is referring to.}
\label{pair_correlation}
\end{figure}

It is obvious that for higher doping density, the antiferromagnetic long range order will be greatly suppressed while the hopping amplitude
between neighboring sites will be enhanced.
At moderate doping level, we can observe that along certain zigzag chains of honeycomb lattice,
hopping amplitude becomes stronger and more homogenous, indicating the existence of possible superconductivity.
\begin{figure}
\centerline{\includegraphics[angle=0,width=8.0cm]{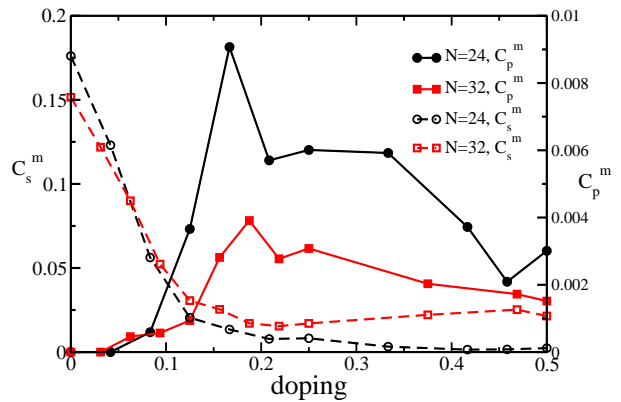}}
\caption{Spin correlation between two furthest separated sites $C^m_s$ and pair correlation
$C^m_p$ between a pair of bonds far separated along zigzag chain of the honeycomb lattice.
We fix $t/J=3$, and tune $\mu$ to control holon doping $n/N$ in the system.
}
\label{order_parameter}
\end{figure}
In order to clarify the nature of the ground state, we measure long range spin-spin correlation $C_s(i,j)$
and pair correlation functions $C_p(i,j)$  defined as
\begin{eqnarray}
C_s(i,j)&=&\langle {\bf S}_i \cdot {\bf S}_{i+{\bf r}} \rangle,\\
C_p(i,j)&=&\langle \tilde{c}^{\dagger}_i\tilde{c}^{\dagger}_{i+{\bf e}} \tilde{c}_j\tilde{c}_{j+{\bf e}} + h.c. \rangle,
\end{eqnarray}
where ${\bf e}$ denotes the vector connecting two nearest neighbors on the honeycomb lattice.
We use spin correlation between furthest separated spins
\begin{eqnarray}
C^{m}_s=1/N\sum_{i}C_s(i,i+{\bf r}_m)
\end{eqnarray}
to detect the strength of long range magnetic order
in the system, where ${\bf r}_m$ depends on size and periodicity of the lattice.
Accordingly, we use pair correlation between third-neighboring bonds
along direction of the unit vector to describe superconductivity in the system, defined as
\begin{eqnarray}
C^{m}_p=1/N\sum_{i}C_p(i,i+{\bf r}_0).
\end{eqnarray}
An illustration of such bond pairs can be found in Fig.~\ref{pair_correlation}.
We plot these two quantities in Fig.~\ref{order_parameter}, as a function of doping,
with $t/J$ fixed to be $3$.
It is quite clear that as soon as the doping density reaches $20\%$,
the long range AF order in the system is greatly suppressed, and superconducting order becomes stable.
We have to point out that sizes of our systems are still relatively small,
and our results is greatly affected by the finite size effect of the system.
To understand physics in thermodynamic limit, one need to calculate much larger system
and do proper size scalings.
GMERA in principle can deal with extremely large lattices, by applying multi-level of renormalization.
The price is limited virtual bond dimension $\chi$ which may
leads to unsatisfactory accuracy.
Calculations for larger lattice is beyond the scale of our current benchmark calculation.
To further explain the superconductivity we observed,
in Fig.~\ref{pair_correlation},
we show pair correlation function $C_p(i,j)$ measured in our system when adequate holes are doped.
For the $N=12$ sites lattice, the pair correlation is much stronger (almost one order larger) for bond pairs
on zigzag chains along one unit-vector shown in the figure. This is due to short period of lattice
along this direction (only 4 sites).
For the $N=24$ sites lattice, this phenomena no longer exists,
and pair correlations are more or less equal with $2\pi/3$ rotational symmetry.
The most important message here is that $C_p(i,j)$ is larger when two bonds
$(i,i+{\bf e})$,$(j,j+{\bf e})$ belong to the same zigzag chain than otherwise,
and remain a considerable number even when these two bonds are furthest separated (limited by lattice size).
Above results hints that superconductivity arises when the system is hole doped,
and current flow along the direction of unit-vector.


\section{IV. Summary}
In this paper we combine the Grassmann algebra with the MERA method based on tensor product state ansatz.
The resulting GMERA method, offers an efficient approach to study 2D strongly correlated electronic system.
To test the validity of GMERA method, benchmark calculations have been performed respectively for
the free-fermion model, the tight binding model and the $t$-$J$  model.
In the case of the $t$-$J$  model, we find that when system is moderately doped, the AF order becomes fully suppressed,
and the superconducting order emerges.

\emph{Acknowledgments.}--- We thank Zheng-Cheng Gu for very stimulating and fruitful discussions. We appreciate Wei Li and Wei Yan for useful discussions and for providing their exact diagonalization and DMRG calculation data for comparison in the benchmark calculation. This work was supported by the State Key Programs of China (Grant No. 2012CB921604), the National Natural Science Foundation of China (Grant Nos. 11304041, 11274069, and 11474064).

\end{document}